\documentclass[twocolumn,trackchanges]{aastex631}
\usepackage{lipsum}

\begin{document}
\makeatletter
\let\frontmatter@title@above=\relax
\makeatother


\newcommand\lsim{\mathrel{\rlap{\lower4pt\hbox{\hskip1pt$\sim$}}
\raise1pt\hbox{$<$}}}
\newcommand\gsim{\mathrel{\rlap{\lower4pt\hbox{\hskip1pt$\sim$}}
\raise1pt\hbox{$>$}}}

\newcommand{\sn}[1]{{\color{blue} Smadar: #1}}
\newcommand{\BH}[1]{{\color{red} BH: #1}}
\newcommand{\IA}[1]{{\color{green} IA: #1}}
\newcommand{\EM}[1]{{\color{orange} EM: #1}}
\newcommand{\AS}[1]{{\color{purple} AS: #1}}

\newcommand{\GAIA}{\textit{Gaia}}
\newcommand{\COSMIC}{\texttt{COSMIC }}
\newcommand{\norm}[1]{\lvert #1 \rvert}


\title{\Large Dynamical Evolution of White Dwarfs in Triples in the Era of \textit{Gaia}}

\author[0000-0003-1247-9349]{Cheyanne Shariat}
\affiliation{Department of Physics and Astronomy, University of California, Los Angeles, Los Angeles, CA 90095, USA}
\affiliation{Mani L. Bhaumik Institute for Theoretical Physics, University of California, Los Angeles, Los Angeles, CA 90095, USA }

\author[0000-0002-9802-9279]{Smadar Naoz}
\affiliation{Department of Physics and Astronomy, University of California, Los Angeles, Los Angeles, CA 90095, USA}
\affiliation{Mani L. Bhaumik Institute for Theoretical Physics, University of California, Los Angeles, Los Angeles, CA 90095, USA }

\author[0000-0001-7840-3502]{Bradley M.S. Hansen}
\affiliation{Department of Physics and Astronomy, University of California, Los Angeles, Los Angeles, CA 90095, USA}
\affiliation{Mani L. Bhaumik Institute for Theoretical Physics, University of California, Los Angeles, Los Angeles, CA 90095, USA }

\author[0000-0002-9751-2664]{Isabel Angelo}
\affiliation{Department of Physics and Astronomy, University of California, Los Angeles, Los Angeles, CA 90095, USA}
\affiliation{Mani L. Bhaumik Institute for Theoretical Physics, University of California, Los Angeles, Los Angeles, CA 90095, USA }

\author[0000-0002-9705-8596]{Erez Michaely}
\affiliation{Department of Physics and Astronomy, University of California, Los Angeles, Los Angeles, CA 90095, USA}
\affiliation{Mani L. Bhaumik Institute for Theoretical Physics, University of California, Los Angeles, Los Angeles, CA 90095, USA }

\author[0000-0001-8220-0548]{Alexander P. Stephan}
\affiliation{Department of Astronomy, The Ohio State University, 4055 McPherson Laboratory, Columbus, OH 43210, USA}
\affiliation{Center for Cosmology and AstroParticle Physics, The Ohio State University, Columbus, OH 43210, USA}


\correspondingauthor{Cheyanne Shariat}
\email{cheyanneshariat@ucla.edu}

\begin{abstract}

The \textit{Gaia} mission has detected many white dwarfs (WDs) in binary and triple configurations, and while observations suggest that triple stellar systems are common in our Galaxy, not much attention was devoted to WDs in triples. For stability reasons, these triples must have hierarchical configurations, i.e., two stars are on a tight orbit (the inner binary), with the third companion on a wider orbit about the inner binary. In such a system, the two orbits torque each other via the eccentric Kozai-Lidov mechanism (EKL), which can alter the orbital configuration of the inner binary. We simulate thousands of triple stellar systems for over $10$~Gyr, tracking gravitational interactions, tides, general relativity, and stellar evolution up to their WD fate. As demonstrated here, three-body dynamics coupled with stellar evolution is a critical channel to form tight WD binaries or merge a WD binary. Amongst these triples, we explore their manifestations as cataclysmic variables, Type Ia supernovae, and gravitational-wave events. The simulated systems are then compared to a sample of WD triples selected from the $\GAIA$ catalog. We find that including the effect of mass loss-induced kicks is crucial for producing a distribution of the inner binary-tertiary separations that is consistent with $\GAIA$ observations.
Lastly, we leverage this consistency to estimate that, at minimum, $30\%$ of solar-type stars in the local $200$ parsecs were born in triples.

\end{abstract}

\section{Introduction} \label{sec:introduction}

The European Space Agency's $\GAIA$ mission \citep{Gaia_Collab} is instrumental in understanding the properties, kinematics, and dynamics of White Dwarfs (WDs). To date, $\GAIA$ has identified $>350,000$ WDs \citep[e.g.,][]{Esteban18,Fusillo19,Fusillo21}, allowing for an unprecedented opportunity to test our theoretical understanding of the dynamical evolution of WDs and their companions. $\GAIA$'s precise measurements of WDs have already led to an abundance of novel insights into WD binary evolution \citep[e.g.,][]{EB18,EB2018b,Cheng20, Ren20,Torres22} and the physical properties (i.e., age, composition, cooling function, etc.) 
of local WDs \citep[e.g.,][]{Cheng19, Blouin20, Chandra20, Tremblay20, Zorotovic20, Blouin21, Torres21, Blouin22, Zorotovic22}. However, the population of observed WD triples from $\GAIA$ has received little attention. 
We aim to leverage this data from the recent $\GAIA$ Data Release 3 to test our understanding of the complex dynamical evolution of triple stellar systems.

Most stars end their lives as WDs.
Interestingly, between 25$\%$-40$\%$
of these WDs reside in a binary or multiple star system \citep[e.g.,][the latter studies focused on the local $25$~parsec]{Raghavan2010,Holberg16,Hollands+18}. 
The multiplicity fraction of WDs is not surprising, given that nearly half of all Sun-like stars are observed to be in binary, or higher order configurations, \citep[e.g.,][]{DM91,Tokovinin1997,Raghavan2010}. In fact, it has recently been suggested that $21\%-36\%$ of wide double WD binaries were once a triple \citep{Heintz22}.
Furthermore, observations suggest that 40$\%$ of older stars have companions (such as WDs), and most ($\geq 70\%$) A and B spectral type stars have one or more companions \citep[e.g., ][]{Raghavan2010, Moe17}. 

WDs in triples and higher order systems are essential to understanding a variety of binary exotica. 
For example, stellar triples and binaries containing at least one WD are key progenitors for Type Ia supernovae
\citetext{\citealp[e.g.,][]{Parthasarathy07, Thompson2011, Katz12, Hamers13, Hamers2018, Toonen18, Michaely2021,MichaelyShara2021,Liu23Review}; \citealp[see][for a review on SNe Ia]{Wang12}}
post-common envelope binaries \citep[e.g.,][]{Toonen13,Zorotovic14,Hernandez22}, and cataclysmic variables \citep[e.g.,][]{Nelemans01, Knigge11,Pala17}. Furthermore, double WD binaries are the most numerous sources of gravitational wave (GW) emission, 
making them primary targets for the prospective Laser Interferometer Space Antenna \citep[\textit{LISA}; e.g.,][]{LISA, Xuan21, Wang21,Seto22}. \textit{LISA} detections of double WDs will provide key insights toward understanding the binary evolution of WDs \citep[e.g.,][]{Korol18} and will allow better constraints on the structure of our Galaxy \citep[e.g.,][]{Breivik20} and the galactic center \citep[e.g.,][]{Wang21,Xuan+23}. 

WDs embedded in triple systems explore a broader range of dynamical behavior, which cannot be observed in binary systems alone. In general, for stability reasons, triple star systems have a hierarchical configuration–two of the stars are in a close binary orbit while the third companion is farther away\footnote{We note that our limitation to a hierarchical configuration is rather conservative. It was shown that deviation from hierarchy can lead to even higher eccentricity excitations rather than immediate instability \citep{Grishin+18QuasiSec, Mushkin+20,Bhaskar+21,Zhang23} }. In hierarchical triples, an interesting dynamical phenomenon gets introduced: the eccentric Kozai-Lidov mechanism 
\citetext{\citealp[EKL;][]{ Kozai1962,Lidov1962}; \citealp[see][for a review]{Naoz2016}}.

EKL-induced effects from a faraway companion star can cause eccentricity and inclination oscillations to the inner binary, which lead to complex dynamical changes to the evolution of the system. High eccentricities caused by EKL, for example, can lead the inner binary to tighten or even merge \citep[e.g.,][]{Naoz2016}. This has been noted to be of special significance in double WD (DWD) binaries, potentially explaining an accelerated rate of observed supernovae \citep[e.g.,][]{Thompson2011}, though the efficiency of this merger channel remains unclear \citep{Prodan13,Hamers13,Hamers2018,Toonen18}. 

Notably, post-main sequence stellar evolution can have a significant effect on the dynamical evolution of binaries and triples \citep[e.g.,][]{PK12,Shappee13,Naoz2016,Stephan16, Petrovich17,Stephan17,Stephan18,Stephan19, Stephan21,Toonen20, Toonen2022, Hamers22, Angelo22,Stegmann22, Kummer23}. 

In this work, we provide a comprehensive investigation of the long-term ($>10$~Gyr) dynamical evolution of stellar triples as
one or more of their components evolve into WDs. We then assess the various outcomes, properties, and signatures of end states that contain WDs. The detailed triple-evolution code includes EKL, general relativity (GR), tides, and post-main sequence evolution, thus allowing us to provide a robust comparison to $\GAIA$'s observed WD population. 

The paper is organized as follows: in Section \ref{sec:methods}, we describe our numerical setup and methodology for the simulations and observations. Section \ref{sec:analysis_and_results} is where we outline and analyze the results of our dynamical simulations. In Section \ref{sec:comp_to_Gaia}, we compare our results to $\GAIA$, taking into account both internal and external perturbations attributed to various physical effects.
Lastly, we discuss our results and provide the major conclusions in Section \ref{sec:conclusions}. Any supplementary equations and figures are provided in the Appendices (Appendix \ref{app: adiabatic}, \ref{kick_app}, \ref{app:param_space}, \ref{app:GAIA_variability}).

\begin{figure*}
\includegraphics[width=1.0
\textwidth]{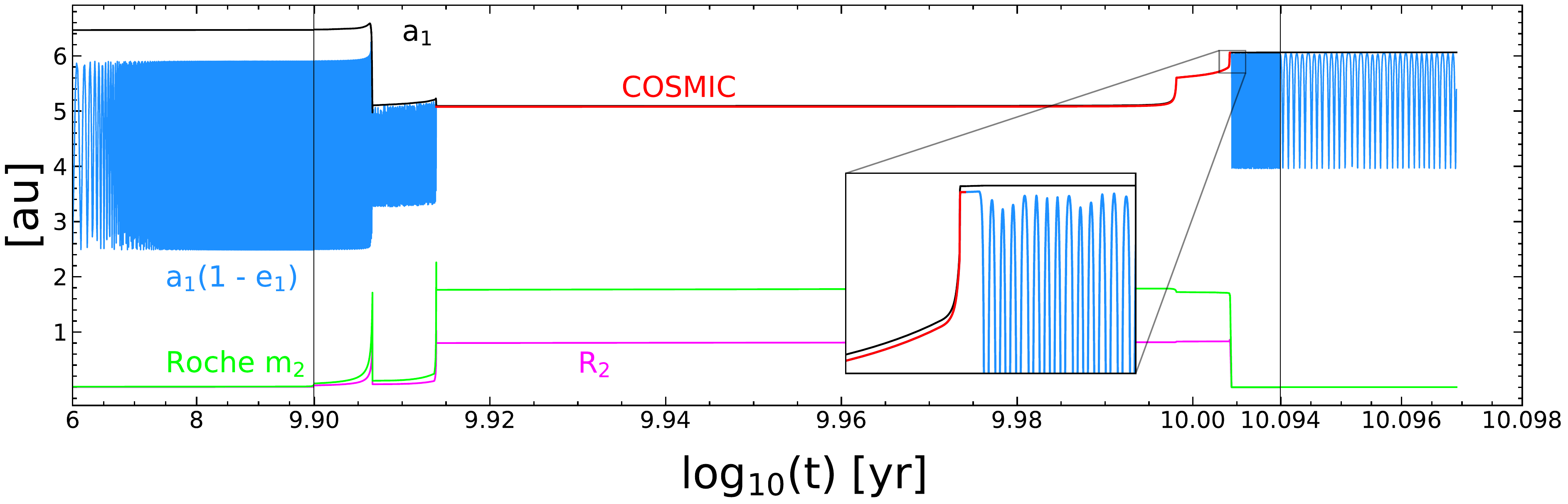}
\figcaption{Time-evolution of the semi-major axis and periastron of the inner binary in a representative triple system. After the $12.5$~Gyr, the system became a double WD inner binary orbited by a low-mass main sequence companion. The oscillating blue line represents the evolution in the triple code, and the red line (labeled COSMIC) shows the binary evolution in \texttt{COSMIC}. As one of the stars in the inner binary evolved into a Giant at  $8.2$~Gyr ($10^{9.91}$~yr), the binary became tidally locked because $e_1 < 0.01$ and $a_1 \leq$ $4$Roche$_2$. We then evolved this binary using \texttt{COSMIC} for $1.9$~Gyr, until the star evolved into a WD, creating a WDMS binary that is no longer tidally locked. After updating the tertiary star's orbital parameters, we then inserted this triple back into the triple evolution code to finish its evolution until $12.5$~Gyr. Initially, this system had $m_1$ = $1.09$~M$_\odot$, $m_2$ = $1.12$~M$_\odot$, $m_3$ = $0.68$~M$_\odot$, $a_1$ = $6.47$~au, $a_2$ = $177.68$~au, $e_1$ = 0.157, $e_2$ = $0.512$, $i_{tot}$ = $55.5^\circ$.\label{time_evol_plot}} 

\end{figure*}

\section{Methodology}\label{sec:methods}
\subsection{Physical Processes and Numerical Setup}\label{sec:num}

Consider a hierarchical triple system of masses $m_1$, $m_2$, on a tight inner orbit, and $m_3$ on a wider orbit.
This system has an inner (outer) semi-major axis $a_1$ ($a_2$), eccentricity $e_1$ ($e_2$), argument of periapsis $\omega_1$ ($\omega_2$) and inclinations with respect to the total angular momentum $i_1$ ($i_2$). Here we use the invariable plane for reference, where the $z$ axis is parallel to the total angular momentum \citep[see][for a full set of equations and definitions]{Naoz2016}.

We solve the equations of motion of the hierarchical triple body system up to the octupole level of approximation; see \citet{Naoz2016} for the full set of equations. We also include general relativistic precession up to the 1st post-Newtonian approximation for the inner and outer orbit \citep[e.g.,][]{Naoz2013GR}. For the mass ratio and scales studied here, these precessions are a sufficient description of the dynamics \citep[e.g.,][]{Naoz2013GR,Lim20,Kuntz22}.  

We adopt the equilibrium tides model for both inner binary members, following \citet{Hut80}, \citet{Eggleton98}, and \citet{Kiseleva98}. This model includes rotational precession, tidal precession, and tidal dissipation \citep[modeled as fixed viscous times of $5$~yr for each star, following][]{Naoz2014}.
Using this tidal description, we can follow the precession of the spin of each star in the inner binary due to the stars' oblateness and tidal torques \citep[e.g.,][]{Naoz2014}. 
We use different tidal models for (radiative) main-sequence and (convective) red-giant stars \citep[e.g.,][]{Zahn97}. 
The switch between tidal models is taking place as a function of stellar type and mass \citep[see][]{Rose+19,Stephan18,Stephan19,Stephan21}. Note that during the WD stage, equilibrium tides are assumed.

Stellar evolution plays an important role in the evolution of triples \citep[e.g.,][]{Naoz2016,Toonen2016,Toonen20,Toonen2022}. Specifically, the mass loss associated with the Asymptotic Giant Branch (AGB) phase can re-trigger the EKL mechanism by changing the mass ratio or by expanding the inner orbit's semi-major axis faster than that of the outer binary \citep{PK12,Shappee13,Michaely2014,Naoz2016,Stephan16,Stephan17}. We thus follow the post-main sequence evolution of stars using the Single Stellar Evolution (\texttt{SSE}) code \citep{SSE}.  See \citet{Naoz2016,Stephan16,Stephan17,Stephan18,Stephan19,Stephan21} and \citet{Angelo22} for a detailed description of the triple with stellar evolution code.

The simulations are run for an upper limit of $12.5$~Gyr but are stopped earlier if the inner binary either (1) crosses the \textrm{Roche} limit or (2) becomes tidally locked. The first condition checks explicitly if the binary is tidally locked.
For the first condition, we define the Roche limit of a star with mass $m_j$, with radius $r_j$ in a binary as \citep[e.g.,][]{Paczynski71,Eggleton83Roche}:
\begin{equation}\label{eq:RL}
    R_{\rm Roche,j}\sim 1.66 \times r_j \left(\frac{m_j}{m_1+m_2}\right)^{-1/3} \ ,
\end{equation}
where $j=1,2$ for the two components of the inner binary. In the second condition, we consider an inner binary to be tidally locked if $e_1< 0.001$ and either $a_1 < 0.1$~au or $a_1\leq 4 R_{\text{Roche}}$.

When the inner binary fulfills either one of these stopping conditions, we follow its
evolution using \texttt{COSMIC} \citet{COSMIC} binary stellar evolution code \citep[procedure similar to][]{Stephan19,Wang21}. \texttt{COSMIC} models the stellar evolution coupled with mass transfer and common envelope and tidal evolution. We keep the default parameters for the code, where the common envelope efficiency is set constant as $\alpha=1.0$. See \citet{COSMIC} for an outline of the other default parameters and their values.

The mass loss during the binary stellar evolution portion is 
modeled as being either adiabatic (slow and isotropic) or impulsive for the tertiary. The adiabatic approximation 
is used when the mass loss of the inner binary during one outer orbit, is much smaller than the total mass of the system.
At that time, we assume that the inner orbit is decoupled from EKL and can only undergo EKL evolution after the binary interaction ends. In this case, we calculate the new orbital parameters of the tertiary star following Appendix \ref{app: adiabatic}. If the mass loss timescale is shorter than $P_2$, we use the impulsive approximation to estimate the new orbital configuration of $m_3$, as outlined in Appendix \ref{kick_app}. See \citet{Lu2019} for an analytic description of the post-kick orbital parameters.

The inner binary is followed in \COSMIC with the following prescription, which depends on characteristic timescales. First, we define the remaining time in the simulation as $t_{\rm remain}$. Second, we consider the quadrupole level of approximation timescale in EKL \citep[e.g.,][]{Antognini2015}, defined by: 
\begin{equation}\label{eq:tEKL}
    t_{\rm EKL} = \frac{16}{30\pi}\frac{m_1+m_2+m_3}{m_3}\frac{P_2^2}{P_1}(1-e_2^2)^{3/2} \ .
\end{equation}
If $t_{\rm EKL} \geq t_{\rm remain}$, we assume the effects of the tertiary are negligible and run  \COSMIC for the remaining time. If $t_{\rm EKL} < t_{\rm remain}$, we evolve the binary in \COSMIC until the mass loss phase has ended. If the binary did not merge during this time, it is put back into triple evolution code so long as it is no longer tidally locked and no longer experiencing Roche Lobe overflow. Figure \ref{time_evol_plot} shows an example of such evolution.

In this Figure, we show the evolution of an inner binary with $m_1=1.09$~M$_\odot$, $m_2=1.12$~M$_\odot$, and outer companion $m_3=0.68$~M$_\odot$. The figure focuses on the semi-major axis (black curve) and pericenter (light blue) evolution as a function of time, starting from $6$~Myr. It also displays the change in the Roche limit (green) and radius (magenta) of $m_2$. The EKL eccentricity oscillations are clearly shown. 
The stars are driven into a tidally locked configuration as the eccentricity approaches 0 and the semi-major axis becomes smaller than four times the Roche limit of $m_2$. When this tidally locked state is reached, the system is evolved using \COSMIC and is assumed to be decoupled from the tertiary. 
The semi-major axis evolution within \COSMIC takes place only due to post-main sequence evolution. The result is two WDs separated by $6$~au. After recalculating the new orbital parameters for $m_3$ using the adiabatic prescription, the system is then evolved again using our triple-body code. The last panel is a zoomed-in on the final $1.3$~Gyr of triple evolution.  

During the \texttt{COSMIC} evolution, we also evolve the tertiary mass ($m_3$) using \texttt{SSE} to follow the change in stellar parameters as the inner binary evolves.

\begin{figure}
\includegraphics[width=0.47
\textwidth]{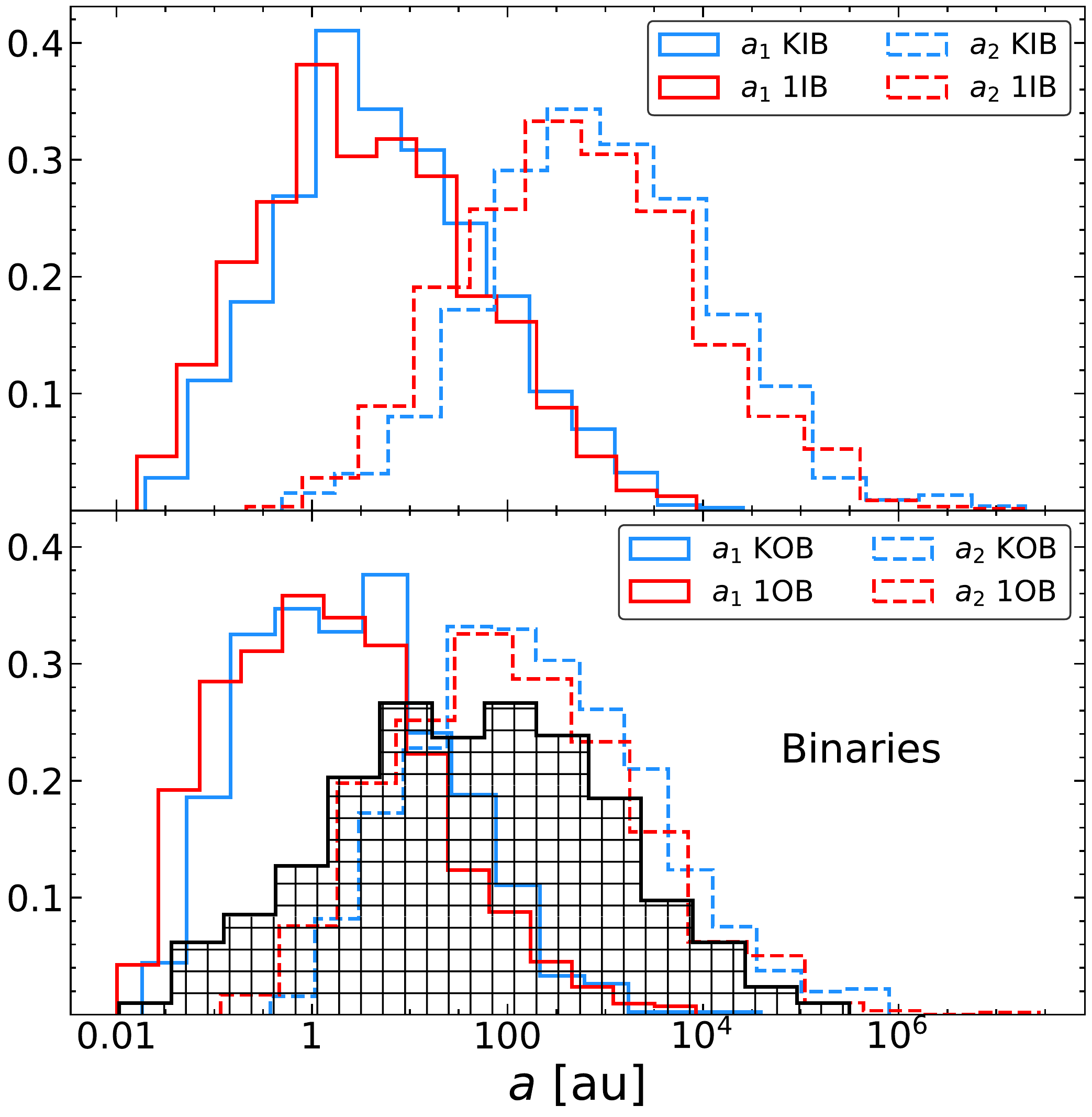}
\caption{ Initial inner (`in') and outer (`out') semi-major axis distributions of the triples for the \textit{1OB}, \textit{KOB} models (top), and \textit{1IB}, \textit{KIB} models (bottom). We also show the Gaussian curve used for our simulated binaries, which is again chosen from the period distribution of \citet{DM91}. This is to illustrate that the OB models had separations samples from this Gaussian curve; for details, see section \ref{sec:ICs} }.\label{fig:ICs} 
\end{figure}

\subsection{Initial Conditions}\label{sec:ICs}

We choose two distinct ways to draw our initial conditions describing two different ansatzes. First, the inner and outer orbital periods are independently chosen from a log-normal period distribution with a mean of $\log(4.8/d)$ and a standard deviation of $\log(2.3/d)$ \citep{DM91}. We then take the larger period to be the outer orbit and the smaller one to be the inner orbit. As the last step, we keep only systems that pass the stability criterion (described in Section \ref{sec:stability}).  This procedure was used in \citet{FT07} and \citet{Naoz2014}, thus allowing us to compare our results with them. We label this set of runs as `IB' (independent binary) because, for unstable systems, both the inner and outer period distributions are re-sampled. 
\citet{Rose+19} and \citet{Stegmann22} showed that the final distribution of periods, (as well as eccentricity), is highly correlated with the initial distribution. Thus, motivated by these results and the observations of \citet{DM91}, we adopt the observed period and eccentricity distributions from \citet{DM91} as the initial distributions of our systems.
The semi-major axis of the inner and outer orbits from this scenario is shown in the top panel of Figure \ref{fig:ICs}.

The second channel draws the outer orbit from the \citet{DM91} period distribution, and for that given outer period, continuously samples the inner orbit until a stable system is formed. This scenario assumes a hierarchy of formation. In other words, the outer orbit may have formed first, thus limiting the parameter space of the stable inner orbit. We note that because we later compare wide outer orbits with a tight inner orbit (in Section \ref{sec:comp_to_Gaia}), we seek to compare both channels to the observations.  We label this set of runs as `OB' (outer binary). The initial orbital separation of OB systems can be seen in  Figure \ref{fig:ICs} bottom panel.


The eccentricity in both cases is chosen from a uniform distribution \citep[consistent with][]{Raghavan2010}, the inclination is chosen from an isotropic distribution (uniform in $\cos i$), and the inner/outer argument of periapsis is chosen from a uniform distribution. The spin angle orbits are chosen from a uniform distribution for all runs. These orbital parameters are also sampled again with the orbital period during the re-sampling phase of both OB and IB models.

Further, we had two different choices for the initial masses. In the first, we chose, for both OB and IB cases, a mass value for $m_1$, $m_2$, and $m_3$ from the Kroupa IMF \citep{Kroupa2001} ranging from $0.8-8$~M$_\odot$ \footnote{Note that a lower mass limit will reduce the fraction of stars that end up as WDs. Because we are interested in the formation of WDs, we keep the minimum mass to be $0.8$~M$_\odot$, which will ensure that a large majority of the binaries will at least have one WD by the end of the simulation time.} (\textit{KOB} and \textit{KIB} runs). This allows us to produce many WD-WD binaries. These models also assume that the mass of the tertiary is independent of the mass of the inner binary, which is consistent with observations of wide binaries \citep{Moe17}.

To compare to \citet{FT07} and \citet{Naoz2014} we also adopt a set of runs where $m_1=1$~M$_\odot$ initially (\textit{1OB} and \textit{1IB}, runs). These runs allow us to produce a robust sample of main-sequence (MS) - WD binaries. The mass ratio $m_2/m_1$ ($m_3/(m_1+m_2$)) for the inner (outer) orbit is chosen from \citet{DM91}, adopting a Gaussian distribution with a mean of $0.23$ and standard deviation of $0.42$. These models assume that the masses between the binaries are correlated. Note that the literature also suggests that various initial mass ratio distributions (IMRDs) are consistent with observations of white dwarf-main sequence binaries \citep{Cojocaru17}. To explore these effects, we run two sets of adjacent simulations. In specific, we assume $m_1 = 1$~M$_\odot$, keep all other orbital elements the same as previously described, and run 250 simulations with a mass ratio selected from $n(q) \propto 1$ (uniform distribution) and 250 with $n(q) \propto q^{-1}$. Here, $q$ is the mass ratio for both the inner and outer binary. After evolving these triples for $12.5$~Gyr, we do not find any qualitative differences between the distributions of the final orbital parameters. We, therefore, omit these models from the paper to avoid clutter. The radii and spin of the stars are directly obtained from \texttt{SSE}.

We sample $1000$ realizations for each of the four models (\textit{1IB}, \textit{KIB}, \textit{1OB}, \textit{KOB}), and provide the statistics in Table \ref{simulations_IC}. A small fraction of the systems continued to run after two weeks of simulation time. Those runs have two categories; in one, they represent double WD (DWD) binaries exhibiting eccentricity and inclination oscillations after $10$~Gyr. For these, we choose the final value for the systems to represent their endpoint (which is longer than $10$~Gyr). These represent $ \sim 1\%$ of the systems. 
The other category represents systems that, after two weeks of running, are still below $12.5$~Gyr time. They slowed down because stellar evolution or tides became important. These triples are omitted from the final sample, representing $<1\%$ of all systems. 

We find that $46\%$ of all triples become tidally locked and $5\%$ cross the Roche limit. After these have been put into \texttt{COSMIC}, $95\%$ of outer parameters were updated according to adiabatic evolution, and only $5\%$ using the kick protocol. Moreover, only $3\%$ needed to be put back into the triple code again, after \texttt{COSMIC}'s binary evolution. 

\subsection{Stability Criteria }\label{sec:stability}

For each set of sampled initial parameters, we require that the initial conditions satisfy dynamical and long-term stability. We adopt the hierarchical criterion $\epsilon$, which describes the pre-factor of the octupole level of approximation \citep[e.g.,][]{Naoz2013sec}
\begin{equation}\label{eq:eps_crit}
    \epsilon = \frac{a_1}{a_2}\frac{e_2}{1-e_2^2} < 0.1 \ .
\end{equation} 
The other stability criterion we use is \citep{MA2001}:
\begin{equation}\label{eq:MA_stability_crit}
    \frac{a_2}{a_1}>2.8 \left(1+\frac{m_3}{m_1+m_2}\right)^\frac{2}{5} \frac{(1+e_2)^\frac{2}{5}}{(1-e_2)^\frac{6}{5}} \left(1-\frac{0.3i}{180^\circ}\right) \ .
\end{equation}
Note that deviation from hierarchy does not necessarily mean an instantaneous breakup of the system or instability. In fact, mildly hierarchical systems can still undergo large eccentricity excitations \citep{Grishin17,Bhaskar21}, and exhibit moderately long-term stability \citep[][]{Mushkin20,Zhang23}. However, here we consider a conservative approach and require only stable and hierarchical systems according to the aforementioned equations.

\begin{deluxetable}{cccclcc}
\tablecaption{Descriptions of Different Simulations\label{simulations_IC}}
\tablehead{
\colhead{Model} &  
\colhead{N} & 
\colhead{$m_1$} & 
\colhead{($m_2$,$m_3$)} &
\colhead{SMA} &
\colhead{Close} &
\colhead{Merged} 
\\
\colhead{} &  
\colhead{Systems} & 
\colhead{} & 
\colhead{} &
\colhead{($a_1$, $a_2$)} &
\colhead{Bin.} &
\colhead{Bin.} 
}
\startdata
\textit{1IB}  & 978 & 1 M$_\odot$     & DM91 & DM91 & 16\%  & 9 \% \\
\textit{1OB} & 929 & 1 M$_\odot$     & DM91 & DM91* &   24\%  & 16\% \\
\textit{KIB}  & 938 & K               & K & DM91  &   6\%   & 56\% \\
\textit{KOB} & 943 & K               & K & DM91*  &     4\%   & 54\%  \\
\textit{1Bin} & 850 & 1 M$_\odot$ & DM91 & DM91  &   27\%   & 13\%  
\\
\enddata
\tablecomments{Close binaries are defined as those with final periods less than 16 days. `DM91*' means that $a_2$ was sampled from DM91, and for that fixed value of $a_2$, $a_1$ was sampled until a stable combination was formed (see Section \ref{sec:ICs}).
}
\end{deluxetable}

\subsection{$\GAIA$ Observations}\label{sec: G_obs}

\begin{figure}
\includegraphics[width=0.46
\textwidth]{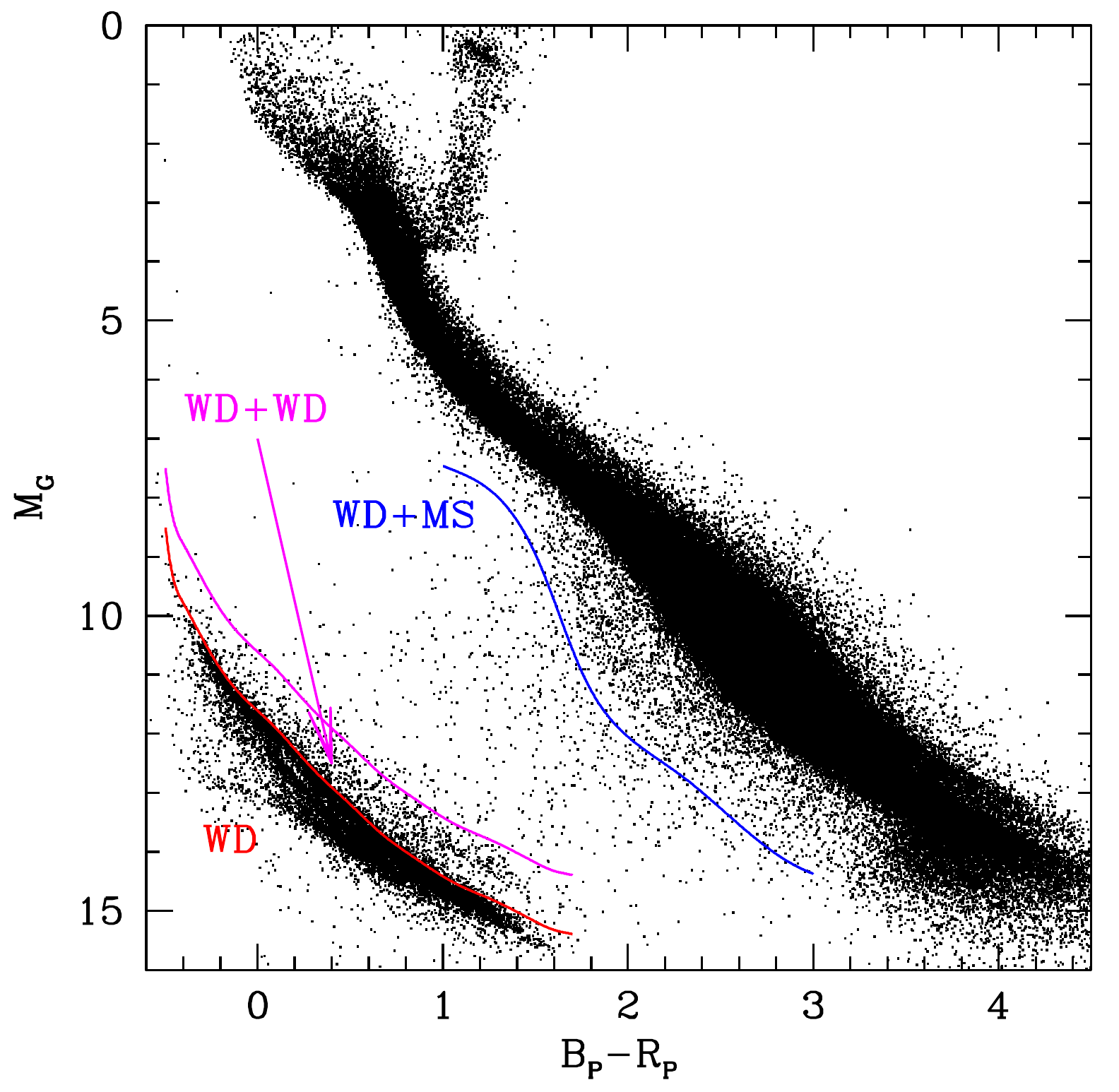}

\figcaption{
We display the flux-magnitude cutoffs used to gather our sample of observed \textit{Gaia} binaries. Each point on this $\GAIA$ HR diagram represents a point source in the \textit{Gaia} DR3 $100$~pc sample. For each source, we the $\GAIA$ magnitude ($M_G$) as a function of the difference between the $\GAIA$ blue ($B_p$) and red ($R_p$) bands. Objects that lie to the left of the red line in Figure~\ref{WDs_HR} are considered to be single WDs. Objects above the red curve and below the pink curve are identified as unresolved double
WD binaries. The range between the magenta and blue lines is the region of WD-main sequence binaries. For more information about the cutoffs, see Section \ref{sec: G_obs}.\label{WDs_HR}}

\end{figure}

The general method of identifying WD triples using $\GAIA$ triples is by first identifying photometric binaries that contain a WD and then matching the binaries with a co-moving companion (of any spectral type). 
Since we are searching for binaries as well, we do not impose any restriction on the \texttt{RUWE} parameter.
After keeping only objects that have photometry for the $G_p$, $B_p$, and $R_p$ \textit{Gaia} bands and have an \texttt{ASTROMETRIC-ERROR} $< 0.9$, we begin our search for WD systems using a $\GAIA$ Hertzsprung–Russell (HR) diagram (Figure \ref{WDs_HR}). 
Objects that lie to the left (bluewards) of the red line in Figure~\ref{WDs_HR} are considered to be single
WDs. This criterion is determined empirically by marking the edge (full width at half maximum) of the number counts as 
a function of color for a fixed $M_G$ bin. Objects that lie above this line are too bright to be a single WD of normal mass at this color and are therefore likely to be an unresolved double
WD -- a photometric binary. In principle, these objects could also be unusually low mass single WDs, but such objects are believed to be the result of close binary evolution anyway, and so still satisfy our criterion. Therefore, we define a second criterion -- the magenta line -- which lies one magnitude brighter than the red line. In between these two criteria, we count objects as photometric double WD binaries. This criterion is similar to the one used in \citet{Inight21}.

WDs may also be in close binaries with main sequence stars as well. For upper main sequence stars, the companion flux would completely overwhelm the WD and these objects are not identifiable by photometry alone. For low-mass main sequence companions, the WD+MS pairs lie between the WD sequence and the main sequence. We, therefore, consider the range between the magenta and blue lines to be the region of WDMS binaries. The blue line was generated empirically, to remove the ``reverse binary sequence'' observed below the main sequence in Figure ~\ref{WDs_HR}. Closer inspection of these objects indicates that they represent blends between foreground stars and distant background objects in crowded regions of the sky, so we exclude this region from the sample selection. Our selection method of WDMS systems is similar to \citet{RM21}.

Any catalog of WDMS systems will be significantly incomplete,
because WD are much fainter than upper MS stars, and will not change the
colors if added together in a photometric binary. To quantitatively assess the level of uncertainty, we construct luminosity functions from the single WD and MS populations and then construct a model WDMS population by randomly sampling from both single populations. For the 26\% of the resulting sample that contain MS stars with $M_G<7$, not even hot white dwarfs are sufficient to generate photometric binaries that fall within our WDMS region. Only 16\% of the model photometric binaries fall within our WDMS region. Furthermore, we have verified that very few WDMS pairs are blue enough to fall within the WDWD photometric cut. Only 0.7\% of our sample fall within this bin -- most importantly, the comparison between the samples defined by our DWD and WDMS cuts imply that the contamination of the DWD sample by the WDMS sample should be only 4\% of the WDMS sample itself.

Despite the low absolute completeness of our WDMS sample, however, the incompleteness is not a major issue in this calculation because our interest is only in comparing the separations of the DWD and WDMS binaries, in relative terms. We are therefore assuming that completeness does not highly correlate with the separation of the outer pair in the triple. See Appendix \ref{app:GAIA_variability} for more detail on our methods of assessing contamination.
  
By comparing these $B_p$ - $R_p$ flux-magnitude values (Figure \ref{WDs_HR}), we identify approximately $60,000$ single WDs, $11,000$ DWDs, and $44,000$ WD\textendash main sequence pairs in the $200$ parsec \textit{Gaia} sample. We note that the other surveys, including those for double WDs \citep[e.g.,][]{Inight21, RM21, Torres22}, are consistent with the estimations in Figure \ref{WDs_HR}.

Once the inner binaries are established, we identify those with a co-moving companion, which would then upgrade the system to a triple. To find these companions, we first require that their distances are consistent with that of the identified primary to within $10\%$ and to within $2\sigma$, where $\sigma$ is the largest of the two distance error bars. For each system, we then calculate the projected separation ($R_{\perp}$), relative projected velocity ($V_{\perp} = V_2 - V_1$), and relative projected angle between the velocities ($\theta$). We require that $\cos(\theta) > 0.99$ to assure that the vectors are pointing in the same directions, and we exclude any object that matched to more than two other objects in the \textit{Gaia} catalog. Lastly, quantifying the  \citet{EB21} velocity criterion, we require the two objects to be in a bound orbit using the empirical criterion: $V_{\perp} < 67,000~km / s (R_{\perp})^{-1.2} $, where $R_{\perp}$ is in au.
This criterion is more relaxed than inclusion criteria based on expected relative velocities for a bound Kepler orbit, but we wish to account for the possibility that the relative motion of the common proper motion pair -- the outer pair of a triple -- may be influenced by partially resolved relative motion within the  binaries -- the inner pair of a triple. The upper limit on the width of the triple is $R_\perp<10^5$~au.

Overall, we find a total of $3,913$ WD binaries, $1,235$ triples with an unresolved WD+WD inner binary, $2,286$ triples with an unresolved WD+MS inner binary, and $87$ triple WDs. In Section \ref{sec:comp_to_Gaia} we compare this observed sample to our simulated population of WD triples.\\ \\ \\ \\

\begin{figure*}
\includegraphics[width=1.0
\textwidth]{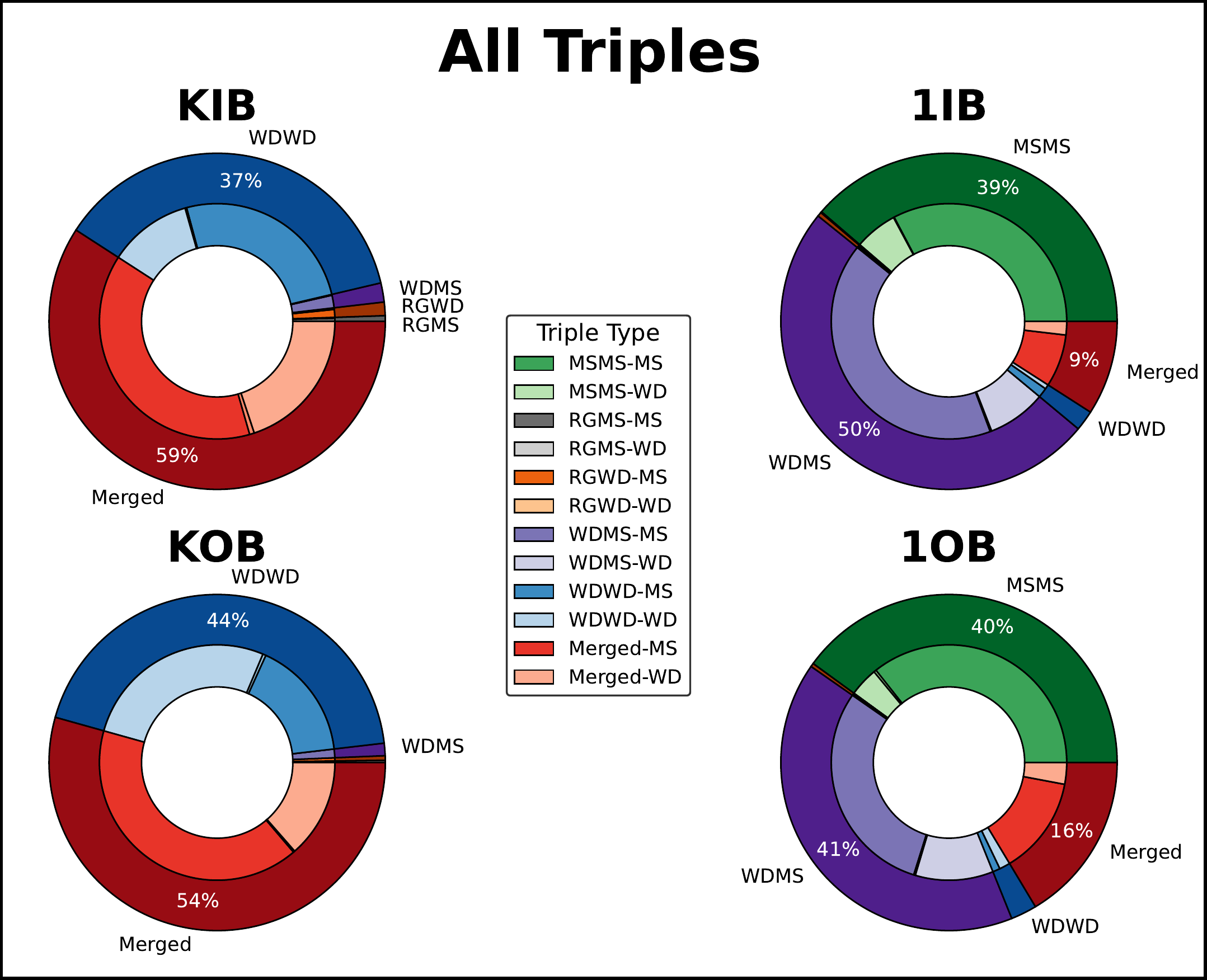}
\includegraphics[width=1.0
\textwidth ]{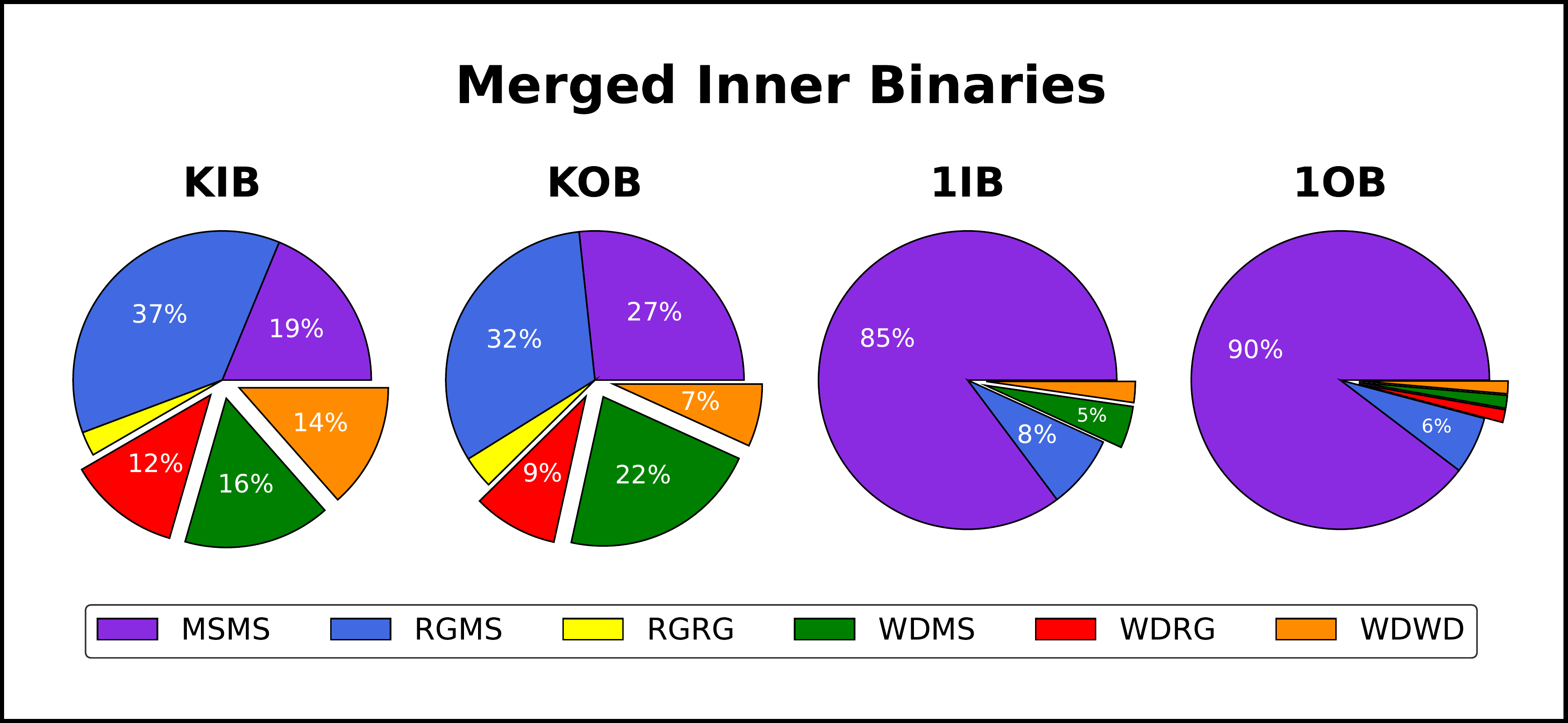}

\figcaption{
The types and fractions
of the different triples (top) and mergers (bottom) produced from the three-body simulations. \textbf{Top:} the outer ring represents the type of the inner binary, and the inner ring specifies the type of the third star. The color corresponding to the type of triples is outlined in the legend, where MS is for main-sequence star, RG is for red giant, and WD is for WD. \textbf{Bottom}: distribution of different stellar types of the inner binary before merger event. The exploded slices correspond to mergers that included at least one WD. 
\label{FinalTypes_pie} }

\end{figure*}

\section{Analysis \& Results}\label{sec:analysis_and_results}

\subsection{Triple Types}\label{sec:Triple_Types}

The specific types of resultant triples from our simulations, along with their fraction is displayed in Figure \ref{FinalTypes_pie}. In this Figure, the first four letters describe the stellar types of the inner binary, and the last two letters describe the type of the tertiary star ($m_3$). For example, WDMS-RG describes a triple with an inner, WD - Main Sequence (WDMS) binary orbited by a Red Giant (RG) tertiary star.

As expected, we find that the Kroupa IMF models (\textit{KIB} \& \textit{KOB}) produce a greater frequency of DWD inner binaries. These models also generate more mergers, mainly due to the accelerated evolution from the larger initial masses. For the $m_{1}=1$ ~M$_\odot$ models (\textit{1IB} \& \textit{1OB}), we find, by design, that the final binaries are mainly WDMS, with a smaller fraction of merged binaries.

\subsection{Outcomes}\label{sec:outcomes}

Figure \ref{FinalTypes_pie} summarizes the outcomes of our simulations. Notably, a sizable fraction of triples merged, especially in the Kroupa IMF models (\textit{KOB} and \textit{KIB}). These were a mix of high eccentricity EKL mergers ($\sim 47\%$), and those that reached a common envelope stage during the post-main sequence evolution ($\sim 53\%$).
We categorize the merged binaries based on the stellar types of the binaries in the bottom panel of Figure \ref{FinalTypes_pie}.

Focusing again on the Kroupa-IMF models, Figure \ref{FinalTypes_pie} shows that $37\%$ and $44\%$ of DWD triples from the \textit{KIB} and \textit{KOB} models, respectively, remained in a triple configuration for the full $12.5$~Gyr. The \textit{KIB} runs assumed an independent choice of the inner binary's initial period, while the \textit{KIB} model chose a fixed sample of the initial outer period before sampling the inner one. See Figure \ref{fig:ICs} for the distribution of initial separations for both \textit{KIB} and \textit{KOB} and Section \ref{sec:ICs} for a description of the different models.

As seen in Figure \ref{FinalTypes_pie}, DWDs are associated with both WD or MS stellar companions. We note that the tight DWD circular binaries ($a_1 \lsim 0.1$~au) are often associated with both WD and MS companions at a wide range of distances (see Appendix \ref{app:param_space} Figure \ref{param_space_cut}). About $7\%$ of all DWD systems exhibit such tight configurations. The rest, as expected, undergo EKL eccentricity oscillations, although, at this point of the evolution ($>10$~Gyr), these are not expected to reach high values. This is because high eccentricity events (due to the octupole-level approximation) would have already taken place. \\ \\ \\

\begin{figure*}
\includegraphics[width=1.0
\textwidth]{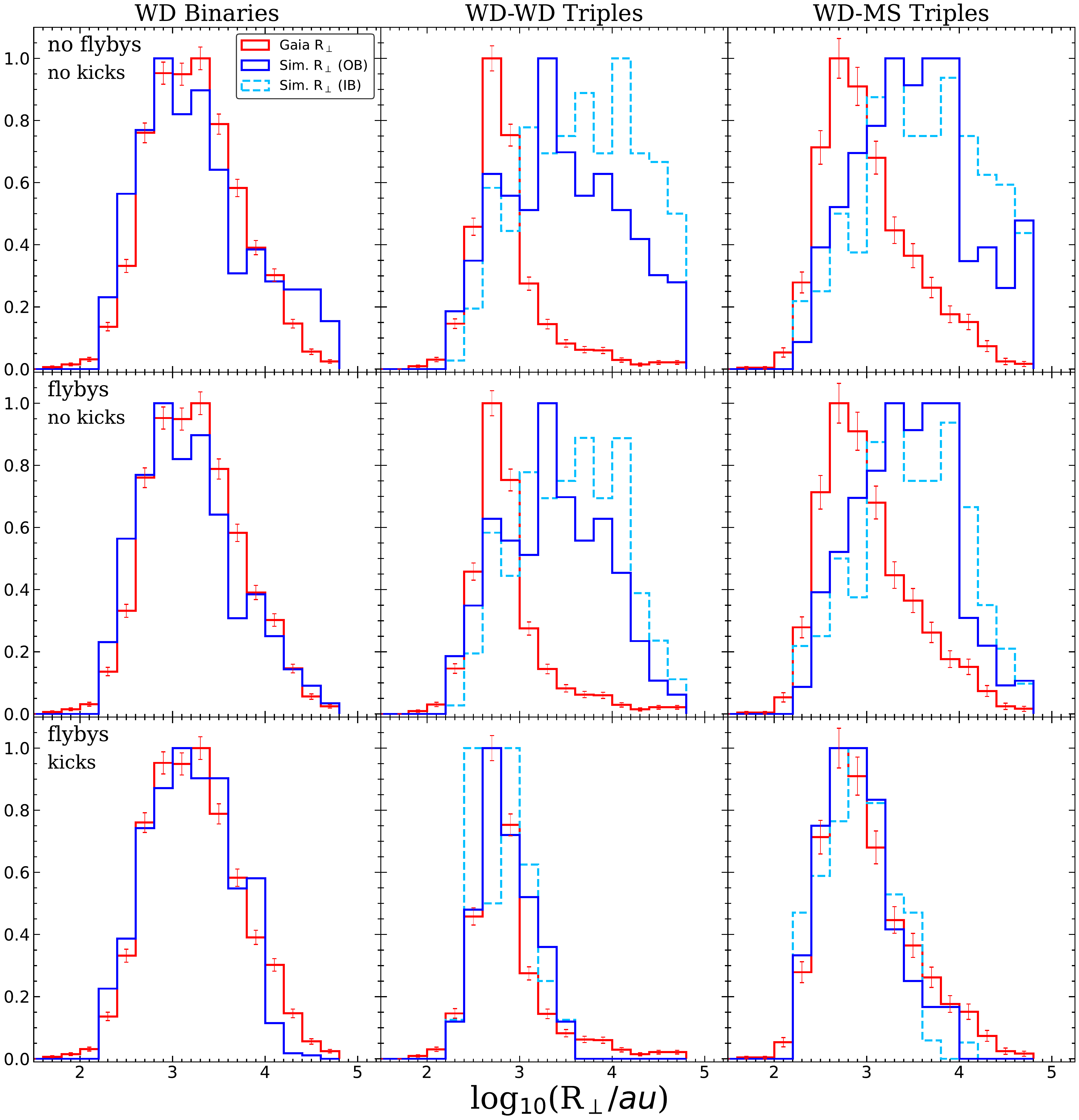}
\figcaption{The Separation Distribution of simulated triples compared to the observed \textit{Gaia} $200$~pc sample.
We compare the projected separations ($R_\perp$) between the inner binary and the tertiary of \textit{Gaia} triples (red) to the final outer semi-major axes $R_\perp$ of the simulated triples. The solid blue curve is the result of our outer binary (OB) models, and the dashed light blue curve is the result of the independent binary (IB) models (Section \ref{sec:ICs}).
We explore the effects of both flyby interactions and mass-loss-induced kicks in changing the separation distributions. The first row shows the raw outcome of simulations, without the inclusion of any perturbative mechanisms. The second row takes into account the effect of flyby interactions in unbinding wide triples \citep[e.g.,][]{Michaely2020}. The bottom row takes into account both flyby interactions and the effect of mass-loss induced kicks in changing the outer separation ($a_2$) of triples \citep[following][]{EB18}. See Section \ref{sec:comp_to_Gaia} for a detailed comparison between the observed and theoretical samples. For information on how flybys and kicks were incorporated into our simulated triple systems, we refer the reader to Sections \ref{subsec:Flybys} and \ref{subsec:Kicks}. Note the agreement between the \textit{Gaia} distribution and the simulation results in the bottom row. \label{sep_obs} }

\end{figure*}
\section{Comparison to \textit{Gaia}}\label{sec:comp_to_Gaia}
\subsection{Setup for the Comparison }\label{sec:filtering_simulated_for_GAIA}
We compare the population of our simulated WD triples to the local $200$~pc $\GAIA$ sample in Figure \ref{sep_obs} (see Section \ref{sec: G_obs} for details on the sample). In Figure \ref{sep_obs}, we compare the distribution of projected separation between the inner orbit and tertiary star ($R_\perp$) of our simulated WD triples to the $R_\perp$ of WD triples from \textit{Gaia}. For the simulated WD systems, we calculate the $R_\perp$ between the inner binary and $m_3$ by relating it to the semi-major axis of the outer orbit ($a_2$) using $R_\perp = a_2/1.10$.
This relation is derived from \citet{Dupuy11}, who show that, for uniform eccentricities, the conversion factor between the semi-major axis of a binary, $a$, and the projected separation, $R_\perp$, can range from $a/R_\perp = 0.75-2.02$ for $1\sigma$ uncertainty, with a median conversion factor of $a/R_\perp = 1.10$. We apply the median conversion factor to convert the simulated $a_2$ to the $R_\perp$ values from $Gaia$. The first, second, and third columns compare the simulated and observed distributions for WD binaries (model \textit{1OB}), WDWD triples, and WD-MS triples respectively.

Specifically, the observed distribution from $\GAIA$ is plotted in red with Poisson error bars. We then compare this distribution to simulations by aggregating the models based on their initial periods. We combine the models with independent choices of the inner binary's initial period (\textit{1IB} and \textit{KIB}) in solid blue, and those where the outer binary was chosen first (\textit{1OB} and \textit{KOB}) in dashed light blue.

For the comparison, we only include simulated triples that exhibit $a_1 < 200$~au and $200$~au $< a_2 < 10^5$~au, to match the observational limitations in the \textit{Gaia} sample (see Section \ref{sec: G_obs}). The restriction on $a_1$ represents the widest inner binary such that it can still be observed as a point source at a distance of $200$~pc with \textit{Gaia}'s angular resolution of $0.43$~arcseconds \citep{Gaia_Collab}. This is the nominal $\GAIA$ angular resolution value, though larger values have been proposed for WD binaries \citep[e.g.][]{Torres22}. We find that such an increase in angular resolution has minimal implications for our comparisons. We investigate the effects of varying angular resolutions with more detail in Appendix \ref{app:GAIA_variability}.
We also note that our method of identifying hierarchical WD triples (see Section \ref{sec: G_obs}) may have a smaller sample of massive DWDs or WDMS binaries with cool WDs. The former is not a major issue because most of our DWD triples have inner binaries with each WD having a mass less than $1$~M$_\odot$. \\ \\

\subsection{Internal and External Perturbations}\label{sec:external_effects}
In order to accurately compare our isolated WD systems to field triples from $\GAIA$, it is crucial that we account for two additional dynamical perturbations: flyby interactions and mass loss-induced kicks \citep[e.g.,][]{Hamers19}. We outline the imprints of both mechanisms in the sections that follow. 

\subsubsection{Flyby Kicks}\label{subsec:Flybys}
The Galactic field is known to be collisional for wide systems with $a > 10^3$~au \citep[e.g.,][]{Kaib2014,Hamers19, Michaely2016, Michaely2019b, Michaely2020}. 
Over the $12.5$~Gyr evolution time, we can expect the wide orbits to experience many weak encounters from field stars. The encounters can serve to (1) ionize wide triples or (2) change the periastron of the outer binary through eccentricity pumping \citep{Michaely2016}. The latter effect can cause a disruption event in the inner binary during the pericenter passage, leading to unstable (or potentially unbound) triples. This effect is more emphasized for wider inner binaries, which exhibit larger loss cone radii.
\citet{Michaely2021} specifically studied the effects of flyby interactions in the field on wide WD triples. They show that, for initially wide triples ($a_2>10^4$~au), there is a non-negligible probability that flyby interactions will destabilize, and potentially unbind, the triple in $\sim 10$~Gyr. 
To account for the effect of ionization of wide triples due to flyby interactions in our sample, we calculate the half-life of the outer binaries following \citet{Bahcall85} and \citet{Michaely2020}. We take the mass of the perturber star to be $0.5~$M$_\odot$, the local stellar number density to be $n_*=0.1$~pc$^{-3}$ \citep{Holmberg00}, and an encounter velocity of $v_{enc}=50$~km~s$^{-1}$. 

Finally, we divide the half-life by the total evolution time to find the probability that the triple will survive after $12.5$~Gyr. If the half-life is greater than the $12.5~$Gyr integration time, we take the probability of survival to be $1$. The height of each bin in the middle row of Figure \ref{sep_obs} is scaled by the probability of survival, which suppresses the survivability of ultra-wide triples. We also expect another fraction of these triples, especially those with wider inner binaries, to become unstable due to effect (2). However, unbinding caused by disruption at the pericenter is a less significant phenomenon that is neglected here \citep{Michaely2020}.

We note that galactic tides can also play a role in disrupting systems in the galactic field with separations larger than $10^4$~au \citep{Kaib2014, Grishin22}. Such effects are neglected because most of these wide binaries already become unbound due to the other two effects.

\subsubsection{Mass-Loss Kicks}\label{subsec:Kicks}
A major internal dynamical effect that could lead to the unbinding or widening of triples is the prospective kicks induced by the post-MS evolution of the inner binary. Previous studies have shown that velocity kicks during post-MS evolution, presumably due to asymmetric mass loss during WD formation, can unbind wide ($a>10^3$~au) systems in the galactic field \citep{Savedoff66,Fellhauer03, Toonen17, EB18}. 

These kicks will vary in magnitude based on the mass of the WD progenitor but will be on the order of $0.75$~kms$^{-1}$ \citep[e.g.,][]{EB18b}. Such an effect was shown to unbind most field binaries with $\log(a/$au$)>3.5$, and lead to a greater correlation with the observed distribution of local $\GAIA$ binaries \citep{EB18}. Following this work on binaries, \cite{Hamers19} investigated the effect of both flyby's and WD kicks on triples. When accounting for both WD kicks and flyby's, up to $50-60\%$ of their triples became unbound in $10$~Gyr. 

Following \citet{EB18}, we assumed that each WD formed in a triple produced a mild, instantaneous kick with velocity $v_{kick}$. The magnitude of this kick was chosen from the Maxwellian distribution 
\begin{equation}\label{eq:Maxwell}
    P(v_{kick}) = \sqrt{\frac{2}{\pi}}\frac{v_{kick}^2}{\sigma_{kick}^3}\exp{\left[-\frac{v_{kick}^2}{2\sigma_{kick}^2}\right ]} \ ,
\end{equation}
which uses a standard deviation $\sigma_{kick}=0.5$~kms$^{-1}$ and peaks at $v_{kick}$=$\sqrt{2}\sigma_{kick}$~$\approx$ $0.75$~kms$^{-1}$ \citep[][]{EB18}. The impact of a natal kick in changing the separation in a system is highly dependent on the direction of the kick and the orbital phase of the companion during the kick.

For each WD in a triple, 
we sample one kick velocity ($v_{kick}$) from the probability distribution in Equation (\ref{eq:Maxwell}). For the chosen kick velocity, we sample the direction of the kick and the eccentric anomaly of the tertiary star $1000$ times, both from a uniform distribution. Then, for each of the $1000$ trials, we calculate the new semi-major axis of the companion ($a_2$) assuming an instantaneous natal kick \citep[see Appendix \ref{kick_app} for the relevant equations, based on ][]{Lu2019}. If more than half of these $a_2$ values led to unbound orbits, we conclude that the kick has ionized the orbit. Otherwise, we choose the median $a_2$ from all samples (that kept a bound orbit) to be the new $a_2$ after the post-WD kick. 

Including the effect of mass loss-induced kicks leads most triples with $\log(a_2/$au$)>3.5$ to become unbound (bottom row of Figure \ref{sep_obs}). Before kicks were applied to the systems (first and second row of Figure \ref{sep_obs}), we find that the WD binaries had consistent distributions with $\GAIA$, while the WD triples did not. This consistency with binaries suggests that our underlying model -- without kicks -- is reasonable. Therefore, the disagreement with the distribution of triples in this panel, and the fact that there is an agreement in the bottom panel, strengthens the argument for the presence of mass-loss kicks during the evolution of triple stellar systems. 

Kicks more strongly affect DWD triples,  because they undergo, at minimum, two kicks during their evolution. The steeper decline in separation distribution for observed DWD triples, compared to WDMS triples, may be attributed to this phenomenon. Namely, the scarcity of wide DWD triples (relative to the number of wide WDMS triples) in our $\GAIA$ field sample may be from their unbinding due to the extra WD kick.


Note that some agreement between the $\GAIA$ sample and our simulations can be also reached by only considering systems where the initial outer-orbit semi-major axis satisfies: $a_2\lsim1250$~au. 

\section{Discussion \& Conclusions}\label{sec:conclusions}

The recent $\GAIA$ Data Release 3 has observed hundreds of thousands of WDs with unprecedented accuracy \citep{Fusillo19,Fusillo21}. Provided that a significant fraction of these WDs have companions \citep{Hollands+18}, $\GAIA$ observations give us a unique opportunity to test our theoretical framework of triple-stellar dynamics on long ($\geq 10$~Gyr) timescales.

In this study, we thoroughly examine stellar three-body systems as they evolve into WD triples. We perform detailed Monte Carlo simulations, where we dynamically evolve thousands of stellar triples for over $10$~Gyr while incorporating, hierarchical three-body secular evolution of the orbits, GR precession, tides, and stellar evolution. Moreover, we track phases of mass loss and the common envelope of the inner binary. %

We leverage $\GAIA$ DR3's data on WDs in triple configurations (see Figure \ref{WDs_HR}) to compare the separation distribution of our simulated WD triples to a $200$~pc sample from $\GAIA$. We find that the DWD-tertiary and WDMS-tertiary separation distributions are consistent with the $\GAIA$ sample if mass-loss kicks are considered. These small kicks ($v_{kick}\sim0.75$~km/s$^{-1}$) may be produced during WD formation \citep{EB18} or otherwise. Their specific origin is kept agnostic in the analysis.

Given the aforementioned agreement, we predict that $\sim 30\%$ of all solar-type stars were born in triples. We derive this estimate by leveraging the statistics from the Kroupa-IMF models (\textit{KIB} and \textit{KOB}). Out of our simulated systems, $55\%$ had  their inner binaries merge within the $12.5$~Gyr. Of the surviving ones, $92.5\%$ ended up in a DWD-tertiary configuration, and lastly, of those systems, $61\%$ became unbound due to mass-loss kicks. Thus, comparing these to the $1235$ DWD-triples candidates that are observed today in the local $200$~pc from $\GAIA$ (see Section \ref{sec: G_obs}), we find  
$7719= 1235/0.16$ DWD triple-progenitors initially.
Given the estimated local stellar density \citep[i.e. $n_*=0.1$~pc$^{-3}$;][]{Holmberg00} we find that $\sim 30\%$ of all WD progenitors were born in triples. 

Moreover, \citet{Heintz22} recently showed that $21\%-36\%$ of wide ($>100$~au) DWDs were likely once triples. They consider the cooling ages of observed WDs in binaries from $\GAIA$ and find shorter cooling ages than would be predicted from the isolated evolution of a WD. This observation is interpreted as evidence of prior mergers or the presence of an unresolved companion. Specifically, these markers would suggest that some fraction of the observed $\GAIA$ DWDs -- as described in Section \ref{sec: G_obs} -- were also once triples. Combining their result with the mentioned fraction of triples estimated from the agreement in Figure \ref{sep_obs} suggests that the fraction of sun-like stars that were born in triples may be even larger, i.e. $\geq 40\%$. 

We note that this fraction is consistent with previous estimates in our local neighborhood \citep[e.g.,][]{Tokovinin1997, Pribulla06}. Moreover, the triple-body evolution also yields many mergers, particularly from the two models, we find that $\sim 22\%$  of all triples lead to a merger containing at least one WD. Specifically, these consist of $10\%$ WDMS mergers, $6\%$ of DWD mergers, and $6\%$ of WDRG mergers. WD merger events may result in (either single or double degenerate) Type Ia supernovae \citep[e.g.,][]{Rosswog09,Raskin09,Hawley12,Hamers13, Hamers2018, Toonen18, Michaely2020, MichaelyShara2021,Liu23Review} and Cataclysmic Variables \citep[CVs; e.g.,][]{Nelemans01,Knigge11, Pala17}. We thus predict $\sim 1698$ such systems within $200$~pc over the last $10$~Gyr. Note that our estimate was derived assuming that all stars were born around the same period of time.

Double WD binaries are also predicted to be the most numerous gravitational wave (GW) sources detectable by \textit{LISA} \citep[e.g.,][]{Marsh11, LISA17}. Specifically, tight or merging DWDs, are expected to be abundant in the mHz \textit{LISA} bands \citep{Korol17,Kupfer18, Burdge19, Li20, Xuan21}. We find the strain curves and signal-to-noise ratio for each of our DWD systems, relative to the \textit{LISA} strain curve. Here, we assume our sources to be $200$~pc. away, the GW emission to be sinusoidal, and the observation time to be four years \citep[the minimum for \textit{LISA}][]{LISA17}. We find that $14\%$ ($\sim 172$ within the local $200$~pc.) of DWDs have a signal-to-noise ratio greater than 5, making them visible in the \textit{LISA} mHz band.


\section{acknowledgements} \label{acknowledgements}
We thank the anonymous referee for constructive feedback on the manuscript. We thank Zeyuan Xuan for useful discussions regarding the gravitational-wave implications of close DWD binaries. C.S. thanks the UCLA Undergraduate Research Fellows Program. S.N. acknowledges the partial support from NASA ATP 80NSSC20K0505 and from NSF-AST 2206428 grant as well as thanks Howard and Astrid Preston for their generous support. 

\clearpage

\appendix
\section{Adiabatic Change in Outer Orbital Parameters}\label{app: adiabatic}

For binaries that evolved in \texttt{COSMIC} for longer than P$_2$, we calculate the new $e_2$ and $a_2$ assuming adiabatic (slow and isotropic) mass loss. Namely, we assume that $e_2$ does not change, and calculate the new $a_2$ from
\begin{equation}\label{adiab_SMA}
    a_{2,f} = \frac{M_{i}}{M_{f}}~a_{2,i},
\end{equation}
where M$_{i}$ is the total mass of the triple before it was inputted into \texttt{COSMIC}, M$_{f}$ is the total final mass of the system, and a$_{2,i}$ is the outer semi-major axis before the binary is inputted into \texttt{COSMIC}. 
We calculate the new mutual inclination (i$_f$) using  
\begin{equation}
    \cos(i_f) = \frac{ G_{tot}^2 - G_{1,f}^2 - G_{2,f}^2}{2G_{1,f}G_{2,f}},
\end{equation}
where G$_{1,f}$ (G$_{2,f}$) is the final orbital angular momentum of the inner (outer) binaries, and G$_{tot}$ is the total orbital angular momentum of the triple \citep[][]{Naoz2016}. For adiabatic mass loss, we assume G$_{2,f}$ to change according to
\begin{equation}
    G_{2,f} = \frac{\mu_{2,f}}{\mu_{2,i}}G_{2,i},
\end{equation}
where $\mu_{2}$ = $(m_{1} + m_{2})m_{3} /(m_{1} + m_{2} + m_{3}) $ is the reduced mass of the outer binary.

\section{Post-Kick Outer Orbital Parameters}\label{kick_app}
For binaries that evolved in \texttt{COSMIC} for less than P$_2$, we calculate the new $e_2$ and $a_2$ by assuming a the inner binary to produce a kick in the outer orbit. We follow the procedure for binaries outlined in \citet{Lu2019}, and use the subscript `1' for the inner orbit, `2' for the outer orbit, and `n' for the post-kick parameters. 

First, we calculate the magnitude of the position vectors of the inner (1) and outer (2) orbits by using 
\begin{equation}
    r_i = a_i\left(1-e_i~cos~E_i\right),  ~ i \in \{1,2\}.
\end{equation}
 Here, the eccentric anomaly ($E_i$) is uniformly sampled for both orbits. The magnitude of the outer orbital velocity is then given by
\begin{equation}
    V_2 = \sqrt{\mu \left( \frac{2}{r_2} - \frac{1}{a_2} \right)},
\end{equation}
where $\mu = k^2(m_3 + m_1 + m_2)$
Then, after defining $\beta \equiv \frac{m_3 + (m_{1,n} + m_{2,n})}{m_3 + (m_1 + m_2)}$, and assuming that the kick velocity ($u_k$) is 0, we apply Equation (10, 19) from \citet{Lu2019} to get the new SMA and eccentricity:
\begin{equation}
    a_{2,n} = a_2\frac{\beta(1-e_2~\cos~E_2~)}{2\beta - (1+e_2~\cos~E_2~)(1+u_k^2+2u_kcos\theta)}
\end{equation}
\begin{equation}
    e_{2,n}^2 = 1 - \frac{r_2^2~V_2^2}{k^2(m_3 + m_{1,n} + m_{2,n})a_{2,n}}.
\end{equation}
Where we have chosen $ r_2 \perp V_2$.

\section{Trends in Orbital Parameters}\label{app:param_space}

\begin{figure*}
\includegraphics[width=0.95\textwidth]
{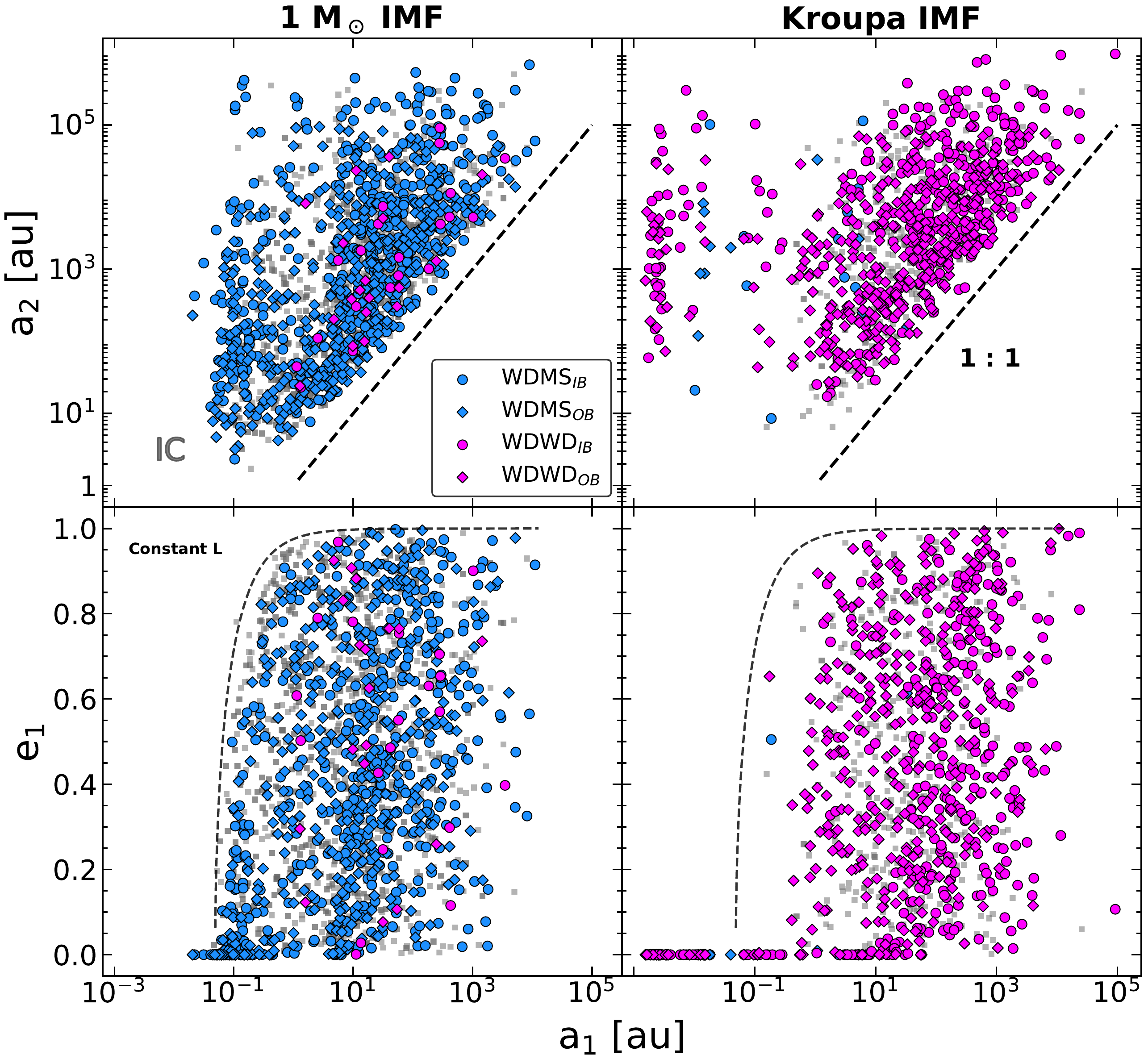}
\figcaption{ The outer orbit separation, ($a_2$, top row) and the inner eccentricity ($e_1$, bottom row), as a function of inner semi-major axis ($a_1$) for all WD triples. The right column depicts the \textit{1IB} and \textit{1OB} runs, for which $m_1=1$~M$_\odot$ initially, while the left panel plots the models with Kroupa IMFs (i.e., \textit{KIB} and \textit{KOB} runs)
We show WDMS binaries in blue and WDWD binaries in magenta. Circles represent binaries from \textit{1IB} and \textit{KIB} models; diamonds represent binaries from the \textit{1OB} and \textit{KOB} models. Grey squares show the initial conditions (IC). The dashed black in the top panel shows the $a_1$ = $a_2$ ($1$:$1$) line. In the bottom row, the dashed black line shows a constant angular momentum curve.\label{param_space_cut}
}
\end{figure*}

We consider systems that kept their triple nature by the end of the run, meaning the inner binary did not merge. Figure \ref{param_space_cut}, depicts the orbital parameters of these triples. In particular, the top panels show the orbital separation of the outer binary ($a_2$) as a function of the inner binary's separation ($a_1$). The bottom panels show the inner orbit's eccentricity ($e_1$) as a function of $a_1$. The grey light markers show the initial conditions. The different colors correspond to the type of inner binary, where WDMS binaries are plotted in blue and WDWD binaries are in magenta. The circular scatter points correspond to triples from the models where the initial periods were chosen independently (\textit{1IB} and \textit{KIB}). The diamond markers refer to triples from the models where the outer orbit's initial period was chosen first (\textit{1OB} and \textit{KOB}). The left column plots only triples from the $1$ M$_\odot$ models (\textit{1IB} and \textit{1OB}) and the right column is for the Kroupa IMF models (\textit{KIB} and \textit{KOB}). See Section \ref{sec:methods} for a complete description of the different models and initial conditions used.

In this figure, the Kroupa-IMF models lead to a greater fraction of tight, circularized WDWD binaries ($a_1<0.1$~au). Interestingly, these close binaries are associated with companions at a wide range of separations.

\section{Impact of Angular Resolution on the $\GAIA$ Sample}\label{app:GAIA_variability}

\begin{figure*}
    \centering
    \includegraphics[width=0.5\textwidth]{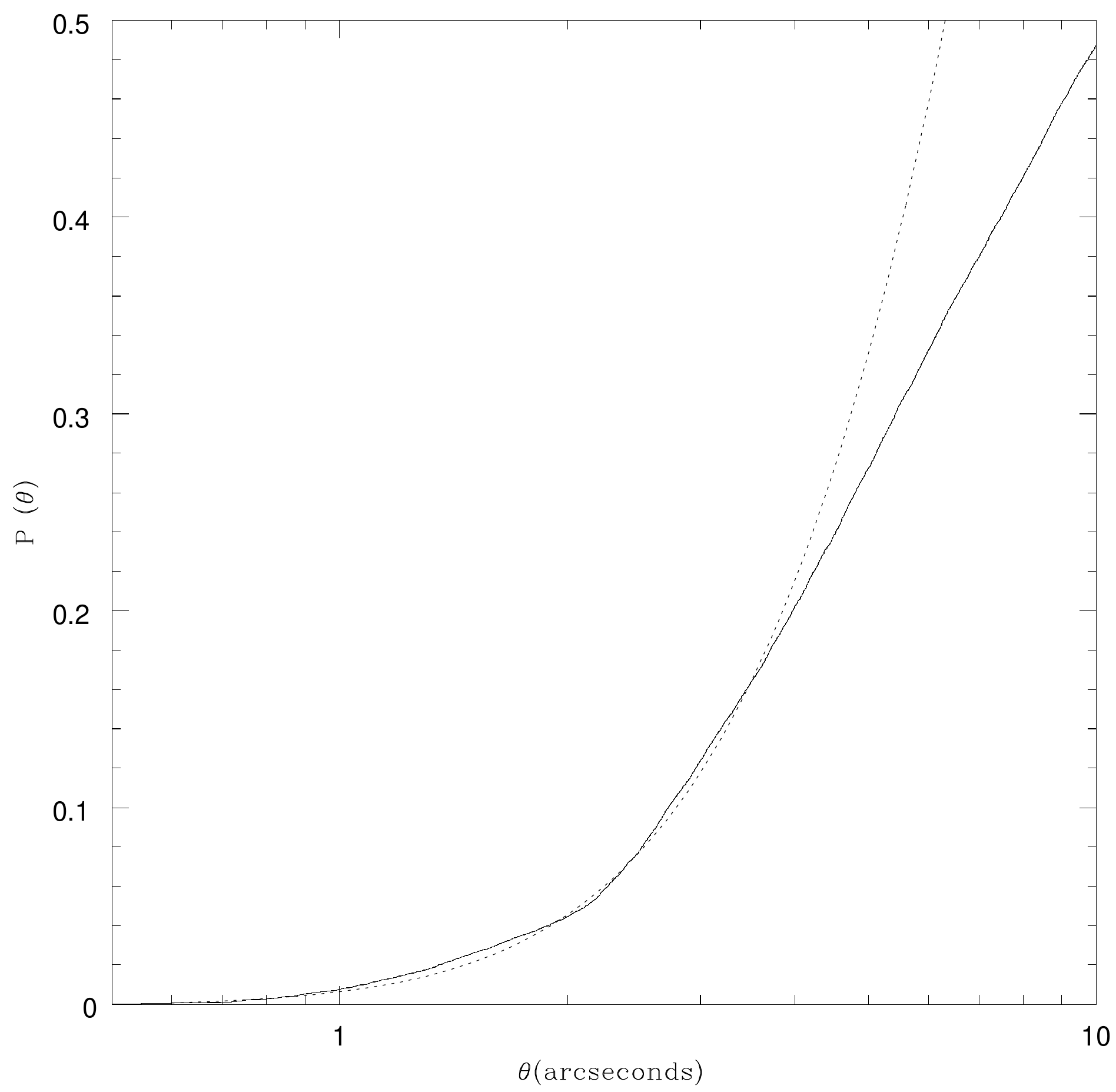}
    \caption{The cumulative distribution function of angular separations for all matched pairs within $200$~parsec in the $\GAIA$ catalog (solid line) compared to the analytic model (dotted) outlined in Equation \ref{eq:sep_gauss}.}
    \label{fig:ang_res}
\end{figure*}

In our analysis, we adopt the nominal $\GAIA$ angular resolution value of $0.43$ arcsec \citep{Gaia_Collab}. Recent studies \citep[e.g.,][]{Torres22} estimate that the angular resolution can reach values closer to $2.5$~arcsec. To assess the impact of the angular resolution on our sample, we investigate the completeness of all photometric pairs within $200$~pc. We test the resolution limit by plotting the cumulative distribution of angular separations for all matched pairs within $200$~parsecs, shown in the solid curve). This curve includes everything, not just WDs. We then compare this distribution to the dotted curve, which is the best-fit Gaussian to our theoretical separation distribution (blue histogram in Figure \ref{sep_obs}). The Gaussian is described by
\begin{equation}\label{eq:sep_gauss}
    n(R)=e^{-0.5(\log(R)-3.15)^2/0.22}.
\end{equation}
Assuming volume completeness out to 200 parsecs, we then convert this separation distribution to an angular resolution distribution (dotted curve in Figure \ref{fig:ang_res}). The dotted curve is an exceptional fit to the solid curve at short separations, suggesting that our angular resolution assumptions are sound.  

As noted above, the specified DWD and WDMS regions identified in Figure \ref{WDs_HR} may be contaminated by other systems. To assess photometric completeness in our sample, we employ a Monte Carlo approach. First, we isolate the single WD and MS, using the cuts specified for our single-star cutoffs (see Section \ref{sec: G_obs}). We then sample both populations and add them together to make an unbiased sample of model WDMS and DWD binaries. We then examine which one of these binaries entered our color-magnitude cuts for the different object types. $25.5\%$ featured an MS star with $M_G<7$, so they are excluded from the sample. $58.3\%$ featured a WDMS binary, but the WD was too faint and it remained rightwards of the blue curve in Figure \ref{WDs_HR}. $15.5\%$ made it into the region between the blue and magenta curves (i.e. our WDMS binaries), $0.66\%$ made it into the DWD region (between magenta and red curves), and $0.036\%$ made it into the single white dwarf region (left of the red curve).
Only $4\%$ of the WDMS sample make it into the DWD region, meaning that about $1760$ of the $11,000$ DWD might be WDMS. As mentioned in Section \ref{sec: G_obs}, our sample does miss WDs around bright MS stars, but this issue is not severe since the contamination fraction is small. \\


\clearpage

\bibliography{references}
\end{document}